\documentclass[twocolumn,preprintnumbers,showpacs,amsmath,amssymb]{revtex4}
\usepackage{bm,epic,eepic}
 \usepackage[english]{babel}
 \usepackage{graphicx}
 \usepackage{pst-plot}
  \usepackage{pstricks}
   \usepackage{pst-coil}

\newcommand{\av}[1]{\langle #1 \rangle}
\newcommand{\BEQ}{\begin{eqnarray}}
\newcommand{\EEQ}{\end{eqnarray}}
\newcommand{\ket}[1]{ |  #1 \rangle}
 \newcommand{\bra}[1]{\langle  #1  |} \renewcommand{\H}{\mathcal{H}} \newcommand{\beq}{\begin{equation}}
\newcommand{\eeq}{\end{equation}}
\newcommand{\Eq}[1]{(\ref{#1})}

\newcommand{\forget}[1]{}

\renewcommand{\vec}{\bm}

\newcommand{\up}{\uparrow}
\newcommand{\down}{\downarrow}

\begin{document}

\title{Local commutativity versus Bell inequality violation  for entangled states and versus non-violation for separable states}

\author{Michael Seevinck}
\email{seevinck@phys.uu.nl}
\author{Jos Uffink}
\email{uffink@phys.uu.nl} \affiliation{Institute of History and
Foundations of Science, Utrecht University, P.O Box 80.000, 3508
TA Utrecht, The Netherlands}

\date{\today}

\begin{abstract}
By introducing a quantitative `degree of commutativity' in terms
of the angle between spin-observables  we present  two tight
quantitative  trade-off relations in the case of two qubits:
First, for entangled states, between the degree of commutativity
of local observables  and the maximal amount of violation of the
Bell inequality:   if both local angles increase from zero to
$\pi/2$ (i.e., the degree of local commutativity decreases), the
maximum violation of the Bell inequality increases. Secondly, a
converse trade-off relation holds for separable states: if both
local angles approach  $\pi/2$  the maximal value obtainable for
the correlations in the Bell inequality decreases and thus the
non-violation increases.
  As expected, the extremes of these
relations are found in the case of anticommuting local
observables where respectively the bounds of $2\sqrt{2}$ and
$\sqrt{2}$ hold for the expectation value of the Bell operator. The
trade-off relations show that non-commmutativity gives ``a more
than classical result" for entangled states, whereas ``a less than
classical result" is obtained for separable states. The
experimental relevance of the trade-off relation for separable
states   is that it provides an experimental test for two qubit
entanglement. Its advantages are twofold: in comparison to
violations of Bell inequalities it is a stronger criterion and in
comparison to entanglement witnesses
 it needs to make less strong assumptions about the  observables implemented in the
experiment.

\end{abstract}
 \maketitle
\section{Introduction}\noindent
In 1964,  John S. Bell \cite{bell64} famously presented an
inequality that holds true for all putative local hidden variable
theories for a pair of spin-1/2 particles but  not in quantum
mechanics. In fact, this inequality is satisfied for every
separable quantum state, but may be violated by any pure entangled
state \cite{gisin}.

It is well-known that in order to achieve such a violation one
must make measurements of  pairs of non-commuting spin-observables
for both particles. It is also well-known (thanks to the work of
Tsirelson \cite{cirelson}) that in order to achieve the maximum
violation allowed by quantum theory, one must choose both pairs of
these local observables to be anticommuting. It is tempting to
introduce a quantitative `degree of commutativity' by means of the
angle between two spin-observables: if their angle is zero, the
observables commute; if their angle is $\pi/2$ they anticommute,
which may thought of as the extreme case of noncommutativity.
Thus one may expect that there is a trade-off relation between the
degrees of local commutativity and the degree of Bell inequality
violation, in the sense that if both local angles increase from 0
towards $\pi/2$ (i.e., the degree of local commutativity
decreases),  the maximum violation of the Bell inequality
increases. It is one of the purposes of this Letter to provide a
quantitative tight expression of  this relation for arbitrary
angles.

It is less well-known that there is also a converse trade-off
relation for separable states. For these states, the bound implied
by the Bell inequality may be reached, but only if at least one of
the pairs of local observables commute, i.e., if at least one of
the angles is zero. It was shown by Roy \cite{roy} (see also
\cite{uffink06}) that if both pairs anticommute, such states can
only reach a bound which is considerably \emph{smaller} than the
bound set by the Bell inequality, namely $\sqrt{2}$ instead  of
$2$. Thus, for separable states there appears to be a trade-off
between local commutativity and Bell inequality
\emph{non}-violation. The quantitative expression  of this
separability inequality was already investigated by
Ref.~\cite{roy} for the special case when the local angles between
the spin observables are equal. It is a second purpose of this
Letter to report an improvement of this result and extend it to
the general case of unequal angles. As in the case of entangled
states mentioned above, the quantitative expression reported will
be tight.

Apart from the pure theoretical interest of these two trade-off
relations, we will show that the last one also has experimental
relevance. This latter trade-off relation is a separability
condition, i.e., it must be obeyed by all separable states, and
consequently, a violation of this trade-off relation is a
sufficient condition for the presence of entanglement.  Indeed,
this separability condition is strictly stronger as a test for
entanglement than the ordinary Bell-CHSH inequality whenever both
pairs of local observables are non-commuting (i.e., for
non-parallel settings).

Furthermore, since the relation is linear in the state $\rho$ it
can be easily formulated as an entanglement witness \cite{witness}
for two qubits in terms of locally measurable observables
\cite{witness3}. It has the advantage, not shared by ordinary
entanglement witnesses \cite{witness,witness2,witness3}, that it
is not necessary that one has exact knowledge about the
observables one is implementing in the experimental procedure.
Thus,  even in the presence of some uncertainty about the
observables measured, the tradeoff relation of this Letter allows
one to use an explicit entanglement criterion nevertheless.

The structure of this Letter is as follows. Before presenting the
trade-off relations in section \ref{tradeoffsection} we will
review some requisite background  in  section \ref{review}. In
section \ref{discussion} we will discuss the import of the
relations obtained.

\section{Bell inequality and local commutativity}\label{review} Consider a bipartite quantum system in the familiar setting of a standard Bell experiment: Two experimenters at distant sites each receive one subsystem and choose to measure  one of  two dichotomous observables: $A$ or $A'$ at the first site, and $B$ or $B'$ at the second.
 We assume that all  observables have the spectrum $\{-1,1\}$.  Define the so-called Bell operator\cite{braunstein}  \beq\label{operator}
  \mathcal{B}:= A\otimes (B +B') +A'\otimes(B-B').
 \eeq
 Since $\av{\mathcal{B}}_\rho:=\mathrm{Tr}[\mathcal{B}\rho]$ is a convex function of the quantum state $\rho$ for the system, its maximum is obtained for pure states.
In fact, Tsirelson\cite{cirelson} already proved that
$\max_{\rho} |\av{\mathcal{B}}_\rho|$  can be attained in a pure
two-qubit state (with associated Hilbert space
$\H=\mathbb{C}^2\otimes \mathbb{C}^2$) and  for spin observables.

In the following it  will thus suffice to consider only qubits
(spin-${1}/{2}$ particles) and the usual traceless spin
observables, e.g.\ $A=\vec{a}\cdot \vec{\sigma}=\sum_i
a_i\sigma_i$, with $\|\vec{a}\| =1$, $i=x,y,z$ and $\sigma_x,
\sigma_y, \sigma_z$ the familiar Pauli spin operators on
$\H=\mathbb{C}^2$.

For the set  $
 \mathcal{D}_{\mbox{sep}}$ of all separable states, i.e., states of
the form $\rho= \rho_1\otimes\rho_2$ on   $\H=\mathbb{C}^2\otimes
\mathbb{C}^2$ or   convex mixtures of such states, the following
Bell
 inequality holds, in the form derived by Clauser, Horne, Shimony and Holt   \cite{chsh}:
 \begin{equation}  \label{chshineq}
     | \langle\mathcal{B}\rangle_\rho|  \leq 2.
 \end{equation}
However, for the set $\mathcal{D}$ of all (possibly entangled)
quantum states, a bound for $ \av{\mathcal{B}}$ is given by the
Tsirelson inequality~\cite{cirelson,landau}:  \beq \label{Tsi}
|\av{\mathcal{B}}_\rho|\leq \sqrt{4+|\av{[A, A'] \otimes
[B,B']}_\rho|}. \eeq

\subsection{Maximal violation requires local anticommutativity}\label{IIA}\noindent
The Tsirelson inequality  \Eq{Tsi} tells us that the only way to
get a violation of the Bell-CHSH inequality  (2) is when both
pairs of local observables are noncommuting: If one of the two
commutators in \Eq{Tsi} is zero  there will be no violation of
(2).  Furthermore, we see from \Eq{Tsi} that in order to maximally
violate inequality \Eq{chshineq} (i.e., to get
$|\av{\mathcal{B}}_\rho|= 2\sqrt{2}$) the following condition must
hold \cite{cirelson, toner}: \BEQ |\av{[A,A']\otimes[B,B']}_\rho|=4.
\label{condition2} \EEQ The  local observables $i[A,A']/2$ and
$i[B,B']/2$ (which are both dichotomous and have their spectra
within $[-1,1]$) must thus be maximally correlated.

However, the condition \Eq{condition2} is only necessary for a
maximal violation, but not sufficient.
 Separable states are also able to obey this condition while  such states never violate the Bell-CHSH inequality.
For example, choose $A=B=\sigma_y$, $A'=B'=\sigma_x$. This gives
$[A,A'] \otimes [B,B'] = - 4 \sigma_z \otimes \sigma_z$. The
condition \Eq{condition2} is then satisfied  in the separable
state $(\ket{\up\up}\bra{\up\up}+
\ket{\down\down}\bra{\down\down})/2$ in the $z$-basis.

Nevertheless, we can infer from \Eq{condition2}  that for maximal
violation the local observables must anticommute, i.e.,
$\{A,A'\}=\{B,B'\}=0$ (a result already obtained in a different
way by Popescu and Rohrlich \cite{popescu}).
  To see this, consider local observables, which are not necessarily anticommuting and note that
  $ i [A, A']/2 = - ( \vec{a} \times
\vec{a'})\cdot \vec{\sigma} $  and analogously $i[B,B]/2= -  (
\vec{b} \times \vec{b'})\cdot \vec{\sigma} $. We thus get
\beq\label{angleres} |\av{[A, A'] \otimes [B',B]}_\rho|=
4|\av{(\vec{a} \times \vec{a'})\cdot\vec{\sigma}\otimes(\vec{b}
\times \vec{b'})\cdot\vec{\sigma}}_\rho       |. \eeq This can
equal 4 only if
$||\,\vec{a}\times\vec{a'}||=||\,\vec{b}\times\vec{b'}||=1$, which
implies that  $\vec{a}\cdot\vec{a'}=0$ and
$\vec{b}\cdot\vec{b'}=0$, since  $\vec{a}$, $\vec{a'}$, $\vec{b}$
and $\vec{b'}$ are unit vectors.

If we denote by $\theta_A$ the angle between observables $A$ and
$A'$ (i.e., $\cos\theta_A=\vec{a}\cdot\vec{a'}$) and analogously
for $\theta_B$, we see that the local observables must thus be
orthogonal: $\theta_A=\theta_B=\pi/2$ (mod $\pi$), or
equivalently, they must anticommute. Thus the condition
\Eq{condition2}   implies that we need locally anticommuting
observables to obtain a maximal violation of the Bell-CHSH
inequality.

 As mentioned in the introduction, local commutativity  (i.e.,  $[A,A']=[B,B']=0$) corresponds to the observables  being  parallel or antiparallel, i.e., $\theta_A =\theta_B =0$ (mod.\ $\pi$), and  local  anticommutativity (i.e.,  $\{A,A'\}=\{B,B'\}=0$) corresponds to the observables  being orthogonal, i.e., $\theta_A, \theta_B=  \pm\pi/2$.
Therefore, in order to obtain any violation at all it is necessary
that the local observables are at some angle to each other, i.e.,
$\theta_A\neq0, ~\theta_B\neq0$, whereas maximal violation is only
possible if the local observables are orthogonal.

This suggests that there exists a quantitative trade-off relation
that expresses exactly how the amount of violation depends on the
local angles $\theta_A,~\theta_B$ between the spin observables. In
other words, we are interested in determining the form of \beq
C(\theta_A, \theta_B) :=  \max_{\rho \in \cal{D}}
|\av{\mathcal{B}}_\rho |\eeq
 In the next section we
 will present such a relation.

However, before doing so, we continue our review for the case of
separable quantum states.  In this case, a more stringent bound on
the  expectation value of the Bell operator is obtained than the
usual bound  of 2.

\subsection{Local anticommutativity and separable states}\label{IIB}\noindent
 Using the quadratic separability inequality of Ref.~\cite{uffink06} for anticommutating observables
($\{A,A'\}=\{B,B'\}=0$) we get for all states in  $
\mathcal{D}_{\mbox{sep}}$:
\begin{align}\label{quadr}
\av{\mathcal{B}}_\rho^2 +\av{\mathcal{B}'}_\rho^2\leq
2[\av{\openone\otimes \openone -&A''\otimes B''}^2_\rho -
\nonumber \\&\av{A''\otimes \openone -\openone\otimes
B''}^2_\rho], \end{align} where $\mathcal{B}'$ is the same as
$\mathcal{B}$ but with the local observables interchanged (i.e.,
$A \leftrightarrow A'$ , $B \leftrightarrow B'$), and where we
have also used the shorthand notation $A''=i[A,A']/2$ and
$B''=i[B,B']/2$. Note that the triple $A,A',A''$ are mutually
anticommuting and can thus be easily extended to form a set of local orthogonal
observables for $\mathbb{C}^2$ (so-called LOO's
\cite{witness2}).

The separability inequality  (\ref{quadr}) provides a very strong
entanglement criterion \cite{uffink06}, but it is here used to
derive a (weaker) separability inequality in terms of the Bell
operator $\mathcal{B}$ for all states in $
\mathcal{D}_{\mbox{sep}}$: \beq\label{Tsisep}
|\av{\mathcal{B}}_\rho| \leq
\sqrt{2(1-\frac{1}{4}|\av{[A,A']}_{\rho_1}|^2)(1-\frac{1}{4}|\av{[B,B']}_{\rho_2}|^2)}.
\eeq Here $\rho_1$ and $\rho_2$ are the reduced single qubit
states that are obtained from $\rho$ by partial tracing over the
other qubit. The inequality \Eq{Tsisep} is the separability
analogue for anticommuting observables of the Tsirelson
inequality \Eq{Tsi}. Note that even in the weakest case  ($
\av{[A,A']}_{\rho_1} = \av{[B,B]}_{\rho_2} =0 $) it implies
$|\av{\mathcal{B}}_\rho|\leq \sqrt{2}$, which  strengthens the
original Bell-CHSH inequality \cite{voetnoot1}. Thus, for
separable states, a reversed effect of the requirement of local
anticommutativity appears than for entangled quantum states.
Indeed, for locally anticommuting observables we deduce from
\Eq{Tsisep} that the maximum value of $\av{\mathcal{B}}_\rho$ is
considerably less than the maximum value of $2$ attainable using
commuting observables. In contrast to entangled states, the
requirement of anticommutivity, which, as we have seen, is
equivalent to local orthogonality of the spin observables,  thus
decreases the maximum expectation value of the Bell operator
$\mathcal{B}$ for separable states.

An interesting question is now: what happens to the maximum
attainable by separable states for locally noncommuting
observables that are not precisely anticommuting? Or put
equivalently, how does this bound depend on the angles between the
local spin observables when the observables are neither parallel
nor orthogonal? From the above one would expect the bound to drop
below the standard bound of $2$ as soon as  the settings are not
parallel or anti-parallel. Just as in the case of general quantum
states it would thus be interesting to get a quantitative
trade-off relation that expresses exactly how the maximum bound
for $\av{\mathcal{B}}_{\rho}$ depends on the local angles of the
spin observables. In other words, we need to establish  \beq
D(\theta_A, \theta_B) :=  \max_{\rho \in
\cal{D}_{\mbox{\scriptsize sep}}} |\av{\mathcal{B}}_\rho|, \eeq
from which we obtain the separability inequality
\beq\label{sepineq} |\av{\mathcal{B}}_\rho|\leq D(\theta_A,
\theta_B), ~~~\rho \in \cal{D}_{\mbox{\scriptsize sep}}. \eeq
 In the following we present such a tight trade-off relation.

\section{Trade-off relations}\label{tradeoffsection} \subsection{General qubit states}\noindent

 \noindent
It was already pointed out by Landau \cite{landau} that inequality
(\ref{Tsi}) is tight, i.e.,  for all choices of the observables,
there exists a state $\rho$ such that : \beq
  \max_{\rho\in \mathcal{D} }
|\langle\mathcal{B}_\rho\rangle|=
 \sqrt{4+|\av{[A, A'] \otimes
[B',B]}_\rho|}. \label{tight}\eeq This maximum is invariant under
local unitary transformations $U\otimes U'$, since Tr$[(U\otimes
U')^\dagger\mathcal{B}(U\otimes U')\rho]=$
Tr$[\mathcal{B}\tilde{\rho}]$ with $\tilde{\rho}=(U\otimes
U')\rho(U\otimes U')^\dagger$. This invariance amounts to a
freedom in the choice of the local reference frames.

Hence, without loss of generality, we can  choose \BEQ \vec{a}
=(1,0,0),~\vec{a'}=( \cos \theta_A ,  \sin \theta_A ,
0),\nonumber\\ \vec{b} =(1,0,0),~\vec{b'}=( \cos \theta_B ,  \sin
\theta_B , 0).\label{choiseobs} \EEQ  For this choice
(\ref{choiseobs}) one has $ i [A, A']/2 =  - \sin \theta_A
~\sigma_z $
   and, analogously,
$i[B,B]/2= - \sin \theta_B  ~\sigma_z $. Hence,   we immediately
obtain \beq\label{Tsiangles}
  \max_{\rho\in \mathcal{D} }
|\langle\mathcal{B}_\rho\rangle| = \sqrt{4+ 4 |\sin \theta_A \sin
\theta_B \av{\sigma_z\otimes\sigma_{z}}_\rho|}. \eeq To obtain a
state independent bound, it remains to be shown that we can choose
$\rho$ such that $|\av{\sigma_z\otimes\sigma_{z}}_\rho|=1$ in
order to conclude  that \beq C(\theta_A , \theta_B)  = \sqrt{4+ 4
|\sin \theta_A \sin \theta_B|}.   \label{C} \eeq

\begin{figure}[h!]
\includegraphics[scale=0.7]{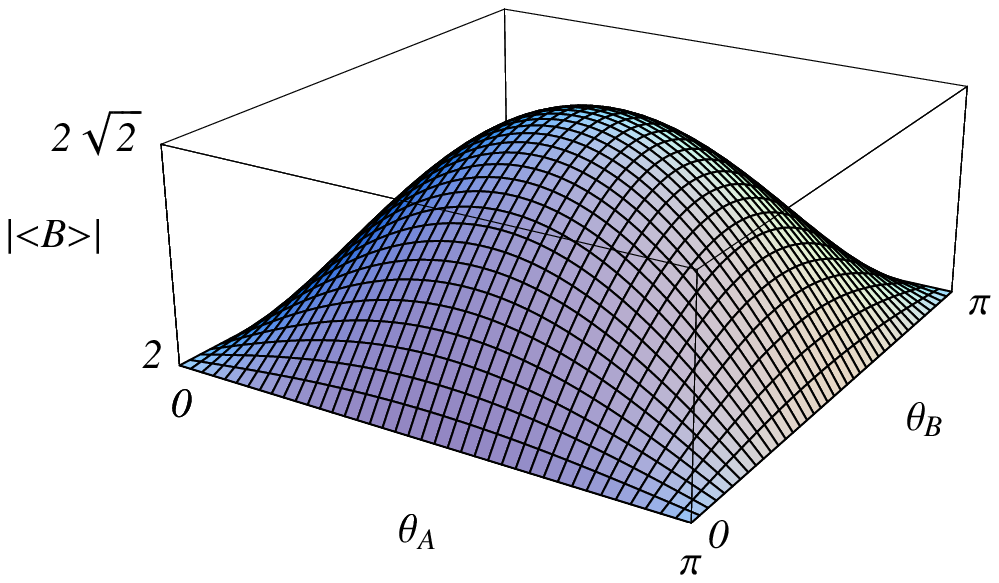}
\caption{Plot of $C(\theta_A,\theta_B)=  \max_{\rho \in \cal{D}}
|\av{\mathcal{B}}_\rho| $ as given in \Eq{C} for $0\leq
\theta_A,\theta_B\leq \pi$.} \label{plaatje_general} \end{figure}

To see that  (\ref{C}) holds,  note that the Bell operator for the
above choice (\ref{choiseobs}) of observables becomes:
\beq\mathcal{B}=
\alpha\ket{\uparrow\uparrow}\bra{\downarrow\downarrow}
+\beta\ket{\uparrow\downarrow}\bra{\downarrow\uparrow} +
\alpha^*\ket{\downarrow\uparrow}\bra{\uparrow\downarrow}
+\beta^*\ket{\downarrow\downarrow}\bra{\uparrow\uparrow}, \eeq
with \BEQ \alpha&=& 1+e^{-i \theta_A}+e^{-i \theta_B}-e^{-i
(\theta_A+\theta_B)}\label{A},\\ \beta&=&1+e^{-i \theta_A}+e^{i
\theta_B}-e^{-i (\theta_A-\theta_B)}\label{B}. \EEQ
 We distinguish two cases: (i) when
$\sin \theta_A \sin \theta_B \geq 0 $  (i.e. when
  $0\leq \theta_A, \theta_B\leq \pi$
 or $\pi \leq \theta_A,\theta_B\leq 2 \pi$),  choose the pure  state $\ket{\phi_{\tau}^+}=\frac{1}{\sqrt{2}}(\ket{\uparrow\uparrow}
+e^{i\tau}\ket{\downarrow\downarrow})$. Then:
\begin{align}
\max_{\tau}\,&\mathrm{Tr}[\mathcal{B}\ket{\phi_{\tau}^+}\bra{\phi_{\tau}^+}]=
\nonumber \max_{\tau}[\mathrm{Re}(\alpha)
\cos\tau+\mathrm{Im}(\alpha)\sin\tau]
\\ &= |\alpha|=
\sqrt{4+4\sin\theta_A\sin\theta_B}.
\end{align}
Similarly,  (ii)  for   $\sin \theta_A \sin\theta_B \leq 0$ (i.e.,
$0\leq \theta_A \leq \pi$, $\pi \leq \theta_B\leq 2 \pi$ or
$\pi\leq \theta_A \leq 2 \pi$, $0 \leq \theta_B \leq \pi$), and
the pure state $
\ket{\psi_\tau^+}=\frac{1}{\sqrt{2}}(\ket{\uparrow\downarrow}
+e^{i\tau}\ket{\downarrow\uparrow})$ we find
\begin{align}
\max_{\tau}\,&\mathrm{Tr}[\mathcal{B}\ket{\psi_\tau^+}\bra{\psi_\tau^+}]=
 \max_{\tau}[\mathrm{Re}(\beta)
\cos\tau+\mathrm{Im}(\beta)\sin\tau]\nonumber \\ & =
|\beta|=\sqrt{4-4\sin\theta_A\sin\theta_B}.
\end{align}
Since
$|\av{\sigma_z\otimes\sigma_{z}}_{\phi_{\tau}^+}|=|\av{\sigma_z\otimes\sigma_{z}}_{\psi_\tau^+}|=1$
we see that the bound in \Eq{C} is saturated. The  shape of the
function $C(\theta_A, \theta_B)$ as determined in (\ref{C})  is
plotted in Figure~\ref{plaatje_general}.

We thus see that  $C(\theta_A, \theta_B)$  becomes greater and
greater when the angles approach orthogonality.  Obviously, for
the extreme cases of parallel and completely orthogonal settings
(i.e., $\theta_A=\theta_B=0$  or $ \pi/2$) we retrieve the results
mentioned in section \ref{IIA}.

If both angles are chosen the same, i.e.,
$\theta_A=\theta_B:=\theta$, \Eq{C} simplifies to
\beq\label{Tsiequalangles}  C(\theta, \theta) = \sqrt{4+ 4 \sin^2
\theta }, \eeq which is plotted in Figure~\ref{grafiek1}.

\subsection{Separable qubit states}
\noindent The set $ \mathcal{D}_{\mbox{\scriptsize sep}}$ of
separable states is closed under local unitary transformations.
Therefore, to find $\max_{\rho \in  \mathcal{D}_{\mbox{\tiny
sep}}} |\av{\mathcal{B}}_\rho|$, we may consider the same choice
of observables as before in \Eq{choiseobs} without loss of
generality. Further, we only have to consider pure states and can
take the state \mbox{$\ket{\Psi}=\ket{\psi_1}\ket{\psi_2}$} with
\mbox{$\ket{\psi_1} =
  \cos \gamma_1 e^{-i \phi_1/2} \ket{\up} +  \sin \gamma_1 e^{i \phi_1 /2}\ket{\down} $}
  and \mbox{$\ket{\psi_2} =  \cos \gamma_2 e^{-i \phi_2/2} \ket{\up} +
  \sin \gamma_2 e^{i \phi_2 /2} \ket{\down}$}.
We then obtain $\av{A}_{\psi_1} =
 \sin 2\gamma_1  \cos\phi_1$,
$\av{A'}_{\psi_1} =
 \sin 2\gamma_1 \cos (\phi_1 -\theta_A)$,  $\av{B}_{{\psi_2}} =
 \sin 2\gamma_2   \cos\phi_2$ and $
\av{B'}_{{\psi_2}}= \sin 2\gamma_2 \cos (\phi_2 -\theta_B).$

\begin{figure}[h!]
\includegraphics[scale=0.7]{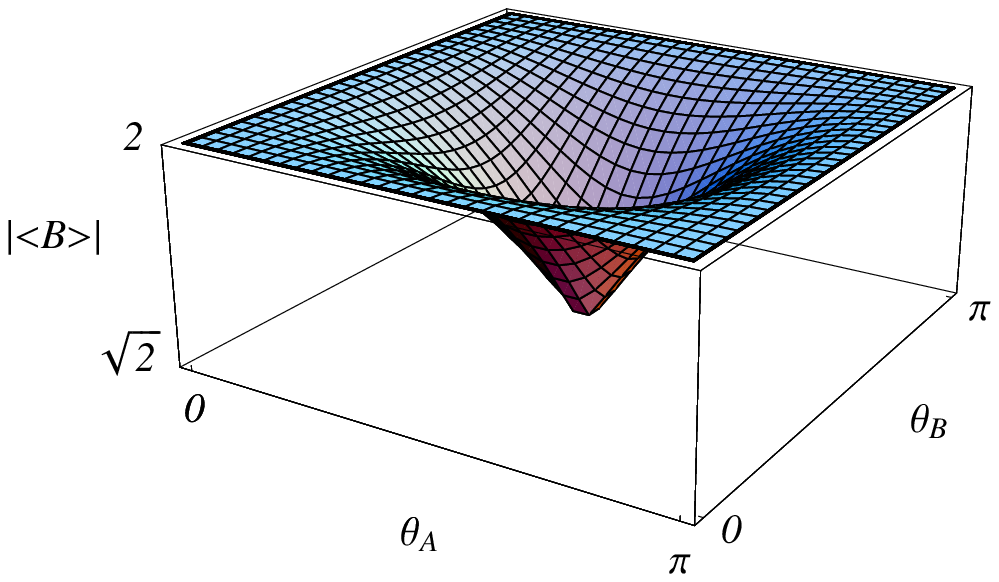} \caption{Plot of
$D(\theta_A,\theta_B) :=\max_{\rho \in \cal{D}_{\mbox{\scriptsize
sep}}} |\av{\mathcal{B}}_\rho|$ as given in \Eq{maxSep} for $0\leq
\theta_A,\theta_B\leq \pi$.} \label{plaatje_separable}
\end{figure}

Since  $\ket{\Psi}$ is separable, we get $\av{A\otimes B}_{\Psi}=\av{A}_{{\psi_1}}\av{B}_{{\psi_2}}$, etc., and the maximal expectation  value of the Bell operator becomes \begin{align} D(&\theta_A, \theta_B) = \max_{{\Psi}}\, \av{\mathcal{B}}_{\Psi } \nonumber\\ &=\max_{\gamma_1, \gamma_2, \phi_1, \phi_2}\sin 2\gamma_1 \sin 2\gamma_2 [\cos \phi_1 (\cos\phi_2 + \cos(\phi_2 -\theta_B))\nonumber\\
 &~~~~+\cos(\phi_1 -\theta_A)(\cos\phi_2 - \cos(\phi_2 -\theta_B) )].\label{sepA} \end{align} This  maximum is attained for $\gamma_1=\gamma_2= \pi/4$ and \Eq{sepA} reduces to:
 \begin{align}
 D(\theta_A, \theta_B)=&\max_{\phi_1, \phi_2} \,\cos \phi_1 (\cos\phi_2 + \cos(\phi_2 -\theta_B)) \nonumber \\
&+\cos(\phi_1 -\theta_A)(\cos\phi_2 - \cos(\phi_2 -\theta_B) ).
\end{align}
A tedious but straightforward calculation yields that the maximum
over $\phi_1$ and $ \phi_2$ is given by \begin{align}\nonumber
D(\theta_A,\theta_B)=& \Big | \mathrm{W}_+ (1 +
\mathrm{X}_{\pm}^{2})^{-1/2} \\ & +
\mathrm{cos}(\mathrm{arctan}(\mathrm{X}_{\pm}) - \theta_A)\,
 \mathrm{W}_-
\Big|,\label{maxSep}
\end{align}
\noindent with
\begin{align}
\mathrm{W}_{\pm}:&=\nonumber (1 + \frac {\mathrm{Z}^{2}}
{\mathrm{sin}^{2}\theta_B\,\mathrm{Y}^{2}})^{-1/2}
    \\&\pm \mathrm{cos}(\mathrm{arctan}(
        {
                \frac {\mathrm{Z}}
            {\mathrm{sin}\theta_B\,\mathrm{Y}}
        } ) + \theta_B),    \\
\mathrm{X}_{\pm} :&=(
\sin\theta_A\cos^2\theta_A\sin^2\theta_B)^{-1}\big(-\cos\theta_A(\cos\theta_B
\nonumber\\&+\cos^2\theta_B +\cos^2\theta_A\sin^2\theta_B)
\pm(\cos^2\theta_A\\
&\times(1+\cos^2\theta_B) (\cos^2\theta_B
+\cos^2\theta_A\sin^2\theta_B))^{1/2}
 \big)\nonumber\\
\mathrm{Y} :&= \mathrm{X}_{\pm}(1- \cos\theta_A + \sin\theta_A),
\\ \mathrm{Z} :&=\mathrm{X}_{\pm}(1 +\cos\theta_B +\cos\theta_A-
\cos\theta_B\,\cos\theta_A) + \nonumber\\&\cos\theta_B\sin\theta_A
-\sin\theta_A, \end{align} \noindent where in $\mathrm{X}_{\pm}$
the $+$ sign is chosen for $-\pi/2\leq\theta_A\leq \pi/2$  and the
$-$ sign is chosen for $\pi/2\leq\theta_A\leq 3\pi/2$ (both modulo
$2\pi$). The function (\ref{maxSep}) is plotted in
Figure~\ref{plaatje_separable}.

\begin{figure}[!h]
\includegraphics[scale=1]{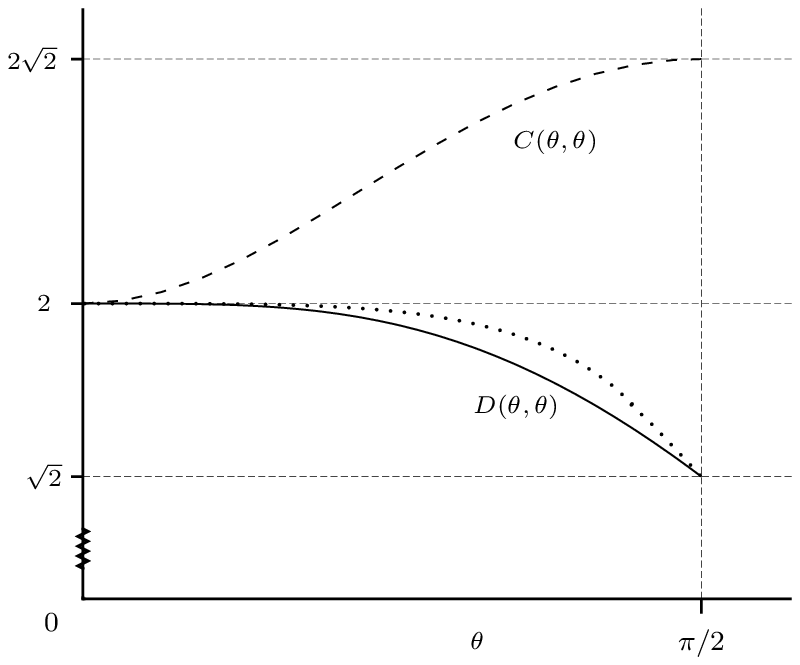}
\caption{Plot of the results  \Eq{Tsiequalangles} (dashed line)
and \Eq{equalsep} (uninterupted line),  and of the bound by Roy
\cite{roy} given in (\ref{royeq}) (dotted line).} \label{grafiek1}
\end{figure}

From this figure we conclude that the maximum of
$|\av{\mathcal{B}}_\rho|$ for separable states  becomes smaller
and smaller when the angles approach orthogonality. For parallel
and completely orthogonal settings we again retrieve the results
of section \ref{IIB}. \noindent  \noindent As a special case, suppose we choose
$\theta_A=\theta_B:=\theta$. Then, (\ref{maxSep}) reduces to the
much simpler expression \begin{align} D(\theta, \theta)=
|\cos\theta| +\sqrt{1+\sin^2\theta}. \label{equalsep}
\end{align}
This result strengthens the bound obtained previously by Roy
\cite{roy} for this special case, which is: \beq
D(\theta, \theta)\leq \left \{ \begin{array}{ll} \sqrt{2}(|\cos\theta| +1),& |\cos \theta|\leq3-2\sqrt{2},\\
1+2\sqrt{|\cos \theta|}-|\cos\theta|,& ~~\mathrm{otherwise}.
\end{array} \right. \label{royeq}
\eeq Both functions are shown in Figure~\ref{grafiek1}.

\section{Discussion}\label{discussion}\noindent
In this letter we have given tight quantitative expressions for
two trade-off relations. Firstly, between the degrees of local
commutativity, as measured by the local angles $\theta_A$ and
$\theta_B$,  and the maximal degree of Bell-CHSH inequality
violation, in the sense that if both local angles increase towards
$\pi/2$ (i.e., the degree of local commutativity decreases),  the
maximum violation of the Bell-CHSH inequality increases. Secondly,
a  converse trade-off relation holds for separable states: if both
local angles increase towards  $\pi/2$, the value attainable for
the expectation of the Bell operator decreases and thus the
\emph{non}-violation of the Bell-CHSH inequality increases. The
extreme cases of these relations are obtained for anticommuting
local observables where the bounds of $2\sqrt{2}$ and $\sqrt{2}$
hold. For the case of equal angles the trade-off relation for
separable states strengthens a previous result of Roy\cite{roy}.

Our results are complementary to the well studied question what
the maximum of the expectation value of the Bell operator is when
evaluated in a certain state (see e.g., \cite{gisin}). Here we have
not focussed on a certain given state, but instead on the
observables chosen, i.e., we asked, independent of the specific
state of the system, what the maximum of the expectation value of
the Bell operator is when using certain local observables. The
answer found shows a diverging trade-off relation for the two
classes of separable and non-separable states.

Indeed,
these two trade-off relations show that local noncommutativity has
two diametrically opposed features: On the one hand, the choice of
locally non-commuting observables is necessary to allow for any
violation of the Bell-CHSH inequality in entangled states (a
``more than classical'' result). On the other hand, this  very
same choice of non-commuting observables  implies  a ``less than
classical'' result  for separable states: For such states the
correlations (in terms of $\av{\mathcal{B}}_\rho$) obey a more
stringent bound than  allowed for in  local hidden variable
theories, i.e. the Bell-CHSH inequality (\ref{chshineq}).

These trade-off relations are useful for experiments aiming to
detect entangled states. They have an experimental advantage above
both Bell inequalities and entanglement witnesses as tests for two qubit
entanglement.
 This will be discussed next.

\begin{figure}[h]
\includegraphics[scale=1]{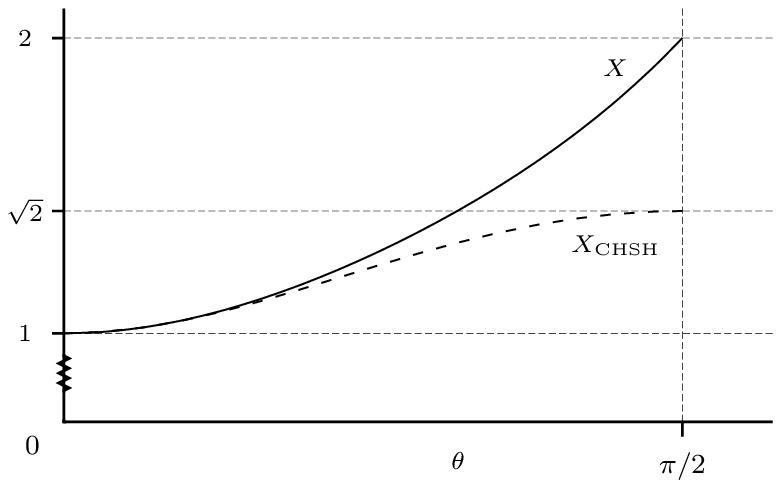}
\caption{Violation factor $X$ (uninterupted line)  and
$X_{\mathrm{CHSH}}$ (dashed line) for
$\theta_A=\theta_B:=\theta$.} \label{grafiek2} \end{figure}

 For comparison to the Bell-CHSH inequality as a test of entanglement,
let us define the 'violation factor' $X$ as the ratio $C(\theta_A,
\theta_B)/D(\theta_A, \theta_B) $, i.e. the maximum correlation
attained by entangled states divided by the maximum correlation
attainable for separable states. In Figure (\ref{grafiek2}) we
have plotted this violation factor $X$  for the special case of
equal angles, cf.\ \Eq{Tsiequalangles}  and \Eq{equalsep} and
compared it to the ratio by which these maximal correlations
violate the  Bell-CHSH inequality \Eq{chshineq}, i.e.
$X_{\mathrm{CHSH}} := C(\theta, \theta)/2$. Figure \ref{grafiek2}
shows  that the  violation factor $X$ is always higher than
$X_{\rm CHSH}$ except when $\theta =0$.  For angles
$\theta\lesssim\pi/4$ these two factors   differ only slightly,
but the violation factor $X_{\mathrm{}}$ increases to $\sqrt{2}$
times the original factor $X_{\mathrm{CHSH}}$ when $\theta$
approaches $\pi/2$. Furthermore note that the factor
$X_{\mathrm{}}$ increases more and more steeply, whereas
$X_{\mathrm{CHSH}}$ increases less and less steeply. For the case
of unequal angles the same features occur, as is evident from
comparing Figures \ref{plaatje_general} and
\ref{plaatje_separable}.

\noindent

Therefore, the comparison of the correlation in entangled states to the maximum
correlation obtainable in separable states yields a stronger test for
entanglement than violations of the Bell-CHSH inequality. Indeed, the violation
factor may reach 2 instead of $\sqrt{2}$. This means that the separability
inequality \Eq{sepineq} allows for detecting more entangled states as well as
for greater noise robustness in detecting entanglement (cf. \cite{uffink06}).
Clearly, the optimal case of this relation obtains when the local observables
are exactly orthogonal to each other. On the other hand, in the case where at
least one of the local pairs of observables are parallel, no improvement upon
the Bell-CHSH inequality  is obtained. But that case is trivial, i.e., no
entangled state can violate  either \Eq{chshineq} or \Eq{sepineq} in that case.

Other criteria for the detection of entanglement than the Bell-CHSH inequality
 have been developed in the
form of entanglement witnesses. In general,  these criteria have
two experimental drawbacks \cite{voetnoot2}: (i) they are usually
designed for the detection of a particular entangled state  and
hence require some a priori knowledge about the state, and  (ii)
they require the implementation of a specific  set of local
observables (e.g., locally orthogonal ones \cite{witness2}). The
separability inequality \Eq{sepineq} compares favorably on these
two points, as we will discuss next.

In real experimental situations one might not be
completely sure about which observables are being measured. For
example, one might not be sure that the local angles are
\emph{exactly} orthogonal in the optimal setup.  However,
  even in such cases, one might be reasonably sure
that the angles are close to 90 degrees, e.g., that
these angles certainly lie within some finite-sized interval $\epsilon$ around 90
degrees. In that case, the bound \Eq{sepineq}  for
separable states would of course be higher than the optimal value
of $\sqrt{2}$ and the increase depends on the size of the
interval specified.  But the trade-off relation presented in this letter
tells us exactly how much higher the bound becomes as a function
of the angles (e.g., $\theta= \pi/2\pm \epsilon$), so one can still obtain a relevant  bound on
$|\av{\mathcal{B}}|$. One can thus still use it as a criterion for
testing entanglement in the presence of some ignorance about  the measured observables.
Entanglement witnesses do not share this
feature since no other  trade-off relations have been obtained (at
least to our knowledge) that quantify how the performance of the witness is changed
when one allows for uncertainty in the observables that feature in
the witness.

Note that for two qubits this result answers the question raised in Ref. \cite{nagata} where it was asked how separability inequalities for orthogonal observables could allow for some uncertainty $\epsilon$ in the orthogonality, i.e., allowing for  $|\{A,A'\}|\leq \epsilon$ (analogous for $B$, $B'$).

A further advantage of the separability inequalities  \Eq{sepineq}
is that they are not state-dependent and are formulated in terms
of locally measurable observables from the start, whereas it is
usually the case (apart from a few simple cases) that
constructions of entanglement witnesses involve some extremization
procedure and are state-dependent. Furthermore, finding the
decomposition of witnesses in terms of a few locally measurable
observables is not always easy \cite{witness3}. However, it must
be said that  choosing the optimal set of observables in the
separability inequalities for detecting a specific state of
course also requires some prior knowledge of this state.

The results presented here only concern the bipartite linear
Bell-type inequality. It might prove useful to look for similar
trade-off relations for nonlinear separability inequalities as
well as for entanglement witnesses. Furthermore, it would be
interesting to extend this analysis to the multipartite Bell-type
inequalities involving two dichotomous observables per party such
as the Werner-Wolff-\.Zukowski-Brukner inequalities \cite{werner}
or the Mermin-type inequalities \cite{mrskab}. For the latter the
situation for local anticommutivity has already been investigated
\cite{roy, nagata, seevinck06}, but for non-commuting observables that are
not anticommuting no results have yet been obtained.


\begin{thebibliography}{99}
\bibitem{bell64} J.S. Bell, Physics {\bf 1}, 195 (1964).
\bibitem{gisin} N. Gisin, Phys. Lett. A {\bf 154}, 201 (1991). N. Gisin and  A. Peres, Phys. Lett. A {\bf 162}, 15 (1992). S.
Popescu and D. Rohrlich, Phys. Lett. A  {\bf 166},  293 (1992).
\bibitem{cirelson} B.S.~Cirel'son, Lett. Math. Phys. {\bf 4}, 93 (1980).
\bibitem{roy} S.M. Roy, Phys. Rev. Lett. {\bf 94}, 010402 (2005).
\bibitem{uffink06} J. Uffink and M. Seevinck, to be published in Phys. Lett A. Quant-ph/0604145 (2006).
\bibitem{witness}M. Horodecki, P. Horodecki and R. Horodecki,Phys. Lett. A {\bf 223}, 1 (1996).
B.M.Terhal, Phys. Lett. A {\bf 271}, 319 (1996); M. Lewenstein, B. Kraus, J.I. Cirac and P. Horodecki, Phys. Rev. A {\bf 62}, 052310 (2000). D.
Bru\ss ~ \emph{et al.}, J. Mod. Opt. {\bf 49}, 1399 (2002).
\bibitem{witness3}O. G\"uhne \emph{et al.},  J. Mod. Opt. {\bf 50}, 1079 (2003). O. G\"uhne \emph{et al.}, Phys. Rev. A. {\bf 66},  062305 (2002). B. M. Terhal, J. Theor. Comput. Sci. {\bf 287}, 313 (2002)
\bibitem{witness2}O. G\"uhne,  M. Mechler, G. T\'oth and P. Adam  Phys. Rev. A \textbf{74}, 010301(R) (2006); S. Yu and N.-L. Liu, Phys. Rev. Lett. \textbf{95} 150504 (2005);  
Zhang, et al., Phys. Rev A. {\bf 76}, 012334 (2007).
\bibitem{braunstein} S.L. Braunstein, A. Mann and M.Revzen, Phys. Rev. Lett. {\bf 68}, 3259 (1992).
\bibitem{chsh} J.F. Clauser, M.A. Horne, A. Shimony and R.A. Holt, Phys. Rev. Lett. {\bf 26}, 880 (1969).
\bibitem{landau}L.J.~Landau,  Phys. Lett. A {\bf 120}, 54 (1987).
\bibitem{toner} B.F. Toner, F. Verstraete, quant-ph/0611001 (2006).
\bibitem{popescu}S. Popescu and D. Rohrlich, Phys. Lett. A {\bf 169}, 411 (1992).
\bibitem{voetnoot1} See  Ref.
\cite{uffink06} for a discussion of how  these separability
inequalities relate to  local realism.
\bibitem{voetnoot2}See also Ref. \cite{enk} where the assumptions  needed in various entanglement
verification procedures are extensively  discussed.
\bibitem{enk}S.J. van Enk, N. L\"utkenhaus and H.J. Kimble, Phys. Rev. A {\bf 75},052318 (2007).
\bibitem{nagata}K. Nagata, M. Koashi and N.Imoto, Phys. Rev. Lett. {\bf 89}, 260401 (2002).
\bibitem{werner} R.F. Werner, M.M. Wolf, Phys. Rev. A {\bf 64}, 032112 (2002). M. \. Zukowski and \v C. Brukner, Phys. Rev. Lett. {\bf 88}, 210401 (2002).
\bibitem{mrskab}N.D. Mermin, Phys. Rev. Lett. {\bf 65}, 1838 (1990); S.M. Roy, V. Singh, Phys. Rev. Lett. {\bf 67}, 2761 (1991); M. Ardehali, Phys. Rev. A {\bf 46}, 5375 (1992); A.V. Belinski\u{\i}, D.N. Klyshko, Phys. Usp. {\bf 36}, 653 (1993).
\bibitem{seevinck06}M. Seevinck and J. Uffink, in preparation.

\end{thebibliography}
\end{document}